\documentstyle[12pt]{article}
\begin{document}
\date{  1996}
\title{The   Partition Function  in the Four-Dimensional Schwarz-Type
  Topological Half-Flat Two-Form  Gravity }

\author{
{\bf Mitsuko Abe }
        \\ \\ \\
     {\it Department of Physics, Tokyo Institute of Technology} \\
     {\it Oh-okayama, Meguro-ku, Tokyo 152, Japan}}
\maketitle
\vskip 2.0cm
\begin{center}
{\bf ABSTRACT}
\end{center}
\abstract{
We derive the  partition functions   
 of the Schwarz-type  four-dimensional topological  
half-flat 2-form gravity model 
on  $K3$-surface or $T^4$ up to on-shell one-loop corrections. 
In this model the bosonic moduli spaces  describe an 
equivalent class of  a  trio of the  Einstein-K\"ahler forms 
( the hyperk\"ahler  forms ).  
The integrand of the partition function 
is represented by the product of some $ \bar 
\partial$-torsions.  
$\bar \partial$-torsion   is the extension of R-torsion for the 
de Rham complex to that for the $ \bar \partial$ 
-complex of a complex analytic manifold. 
\thispagestyle{empty}
\newpage
{\bf  Introduction}
\par 
Recently, Witten gave  some gravitational versions of topological 
quantum field theories~ \cite{witten2}.  These theories are 
 important as the effective theories of the ordinal gravity theories. 
 For example, he pointed out the  relation  between   
the two-dimensional topological gravity models   and the string 
theory~~ \cite{witten2}. He also gave four-dimensional topological 
Yang-Mills  theories with $N=4$ or $N=2$ twisted supersymmetry  whose 
partition function  is conjectured to satisfy S-duality~\cite{witten3}. 
These  observations  seem to be necessary to obtain the 
non-perturbative effects  of the string theories  or   the gravity theories.
\par
Since the work of Witten, there have been several attempts to 
construct  four-dimensional topological gravity theories  over 
 different kind of the gravitational moduli 
spaces~~ \cite{perry}-\cite{kunitomo}.  
\par 
There  two types of models have been proposed 
 for the  four-dimensional half-flat 2-form topological gravity. 
(A) Witten-type topological gravity model, which was given by Kunitomo~~
\cite{kunitomo} and 
(B) Schwarz-type \cite{schwarz}
 topological gravity model \cite{lee,abe}. 
The bases of their formalism are given by ref. \cite{capovilla,samuel,horowitz}. The interesting relation  between the half-flat gravity 
and 2-dim. conformal field theory is investigated by Park~~\cite{park}.
In the previous paper we showed that 
by taking the suitable gauge fixing condition and the limit of the 
coupling constant  for (B),  the bosonic part of the    moduli spaces of 
(B) coincides with that of (A). 
These moduli spaces  are  those   of  the Einstein K\"ahlerian  
manifolds with  vanishing   real first Chern class~\cite{abe}.
\par
The purpose  of  this paper is to 
examine  the  partition function 
of   the  Schwarz-type   model 
and to know whether its 
integrand  is represented by the R-torsion or the 
$\bar \partial $-torsion \cite{ray}
\footnote{There  have been  already several results 
 about  renormalization of non-abelian  BF-type  model  
in the Landau gauge or the covariant gauge\cite{blau,nakamichi}. 
Our gauge fixing condition is different from  them. 
Ours  are introduced  to specify 2-form fields as the 
pre-metric field ( or as the hyperk\"ahler forms at last). 
Thus  our  model is  different from their model.}
  up to on-shell one-loop corrections.
We concentrate our attention for  $K3$-surface and $T^4$ cases 
only. Thus the action reduces to the Abelian BF-type
\footnote{"BF-type" means  that the action consists of 
Lie algebra valued field strength $F$ -field and  Lie algebra valued 2-form 
$B$.}  model with the special gauge fixing conditions, 
which we call the diffeo. BF-type model.  
Their moduli spaces are identified with the 
deformation of a trio of the Einstein-K\"ahler  forms 
( the hyperk\"ahler forms )  which  is  
related to the Plebansky's heavenly equations~~\cite{plebanski}. 
\par
The extension of the algebraic curve with Einstein metric to the 
four dimensional cases may be the algebraic surfaces with Einstein metrics. 
$T^4$ and $K3$ surface belong to the algebraic surfaces and we regard   
these models as the simple examples that treat the algebraic surfaces.
\par
As  another aspect, 
there have been discovered rich type of non-compact gravitational 
instantons (i.e. ALE ~~\cite{eguchi} or ALF ~~\cite{hawking}) 
which satisfies these equations.
In this paper, we will treat  the compact manifolds  only.  
In the near future, we will extend our investigation to non-compact case.
\par Furthermore,  the $N=1$  or $N=2$ 
supersymmetric extensions of chiral 2-form gravity model
are given by \cite{capovilla,kunitomo2}. The supersymmetric 
extension of our model by using them  would be a  toy model  
 for researching the moduli spaces of  the compactified string theory . 
\newpage
{\bf   The four-dimensional topological  half-flat  
 2-form  gravity models}
\par
The Schwarz-type  2-form gravity action 
 reduces to the Abelian BF-type action\cite{abe}
with our gauge fixing condition which we introduce later:
\begin{equation} 
S_0^{\rm red.}= \lim_{\alpha \rightarrow 0} {1 \over \alpha}
  \int_{M_4}\Sigma^k \wedge  d\pi_k,   
\end{equation}
where $M_4$ is a four dimensional manifold and  
we consider only  $M_4=K3$ or $T^4$.
$\alpha$ is a dimensionless parameter and $k=1,2,3$.
We restrict this model as the gravitational one. 
\par 
$\pi_k$ and $\Sigma_k$ are   
 self-adjoint 1-form  and  2-form fields. 
 Their deformations are   $(2K\oplus O)^\ast$ valued 1-form and 
$(2K\oplus O)^\ast$ valued 2-form 
\footnote    
 {In this action, we  use  that   
$(2K\oplus O ) \otimes \wedge^3 \cong 
(2K\oplus O )^\ast \otimes \wedge^1
\cong (2K\oplus O ) \otimes \wedge^1$.  It comes from the fact that 
$K$ is trivial on $K3$ and $T^4$.},
  where $O$ is a  trivial bundle  and $K$ is a canonical line bundle on $M_4$.  They are both independent on the spin connections     
\footnote 
{We slightly change the model in \cite{abe} such as  $\pi_i$  is independent on  the spin connections.}.
This action  is  proper for $M_4= K3$ or $ T^4$ 
 with our gauge fixing conditions. 
The reason is that  
 the canonical  bundles $K$ and 
 $P_{U(1)}$ (the principal $U(1)$ bundle
that  comes  from  $P_{U(2)}$  of oriented orthonormal frames )
  are trivial on   $K3$-surface  or $T^4$.
 Thus the  reductions of $P_{U(2)}$ of    oriented orthonormal frames 
 are possible    when they have  Einstein-Kahler metrics on them. 
Therefore  the chiral part of the local Lorenz symmetry of
$U(1)_R$ disappears in the definition of the moduli space.
These manifolds  are called hyper-K\"ahlerian \cite{besse}.
Thus fundamental fields in this model  are  a trio of  2-form 
$\Sigma^k=\Sigma^k_{\mu\nu} dx^\mu  \wedge dx^\nu$. 
The symmetries of this model  are the diffeomorphism and the  
redundant symmetry (p-form symmetry)
 because the  action is  an Abelian BF-type one. 
\begin{equation}
\delta_B \Sigma^k = \delta _{\rm diffeo} \Sigma^k+ d\phi^k,~
\delta_B \pi^k = \delta _{\rm diffeo} \pi^k+ d\Pi^k,
\end{equation} 
where $\phi^k$ is a fermionic field with $(2K \oplus O)^\ast$ valued 1-form 
and $\Pi^K$ is a fermionic field  with $(2K \oplus O)^\ast$ valued 0-form.
The moduli space of this model depends on  the gauge 
fixing conditions which we take. 
\par 
We  set the following gauge fixing conditions to fix the 
p-form symmetries modulo  diffeomorphism 
and make $\Sigma^k$ as the pre-metric fields.  
\begin{equation}
    {}^{t.f.} \Sigma^i \wedge \Sigma^j 
    \equiv \Sigma^{(i} \wedge \Sigma^{j)} 
    - {1 \over 3} \delta_{ij} \Sigma^k \wedge \Sigma_k = 0.
\label{eq:three}
\end{equation}
From this,  $\Sigma^k$    
 comes from a   vierbein  $e^a= e^a_\mu dx^\mu$~\cite{capovilla} :
\begin{equation}
 \Sigma^k(e)  = - \bar \eta^k_{ab} e^a \wedge e^b \propto  
g_{\alpha \bar \beta}  J^{k \bar \beta }_{\bar \gamma} dz
^\alpha \wedge d\bar z^ {\bar  \gamma}, 
\label{eq:four}
\end{equation} 
where $\bar \eta^i_{ab}$ is the t'Hooft's $\eta$-symbol 
\cite{t'hooft}.
$\{ \Sigma^k(e) \}$ has 13 degrees of freedom. 
$\{ J^k \}$ represents   three  almost complex structures 
which  satisfies the  quaternionic relations and 
$g_{\alpha \bar \beta}$ is an hermite  symmetric metric. 
One of equation of motion is  given by 
\begin{equation}  
d\Sigma^k=0\ .
                                        \label{eq:five}
\end{equation}
Eq. (3) and eq. (5)  
get  $\Sigma^k$ to be   the Einstein-K\"ahler forms (hyper-K\"ahler forms). 
\\
\par 
The  moduli space is the equivalent class of  a trio of the   
Einstein-K\"ahler forms  (the hyperk\"ahler forms) $\{\Sigma^k(e)\}$~
\cite{abe}.
\newpage
{\bf  The gauge fixing conditions  for the BRST quantization      }
\par  
We discuss about the partition function of the Schwarz-type 
model on $K3$ or $T^4$.  The BRST symmetry
\footnote{We would not do the off-shell extensions of these 
symmetries in this paper since the changes caused by the 
off-shell extensions would only work as the higher order terms. }
 and the action   is  given by \cite{abe}.  
This action is invariant under the diffeomorphism  transformations and   
the  p-form symmetry.  These transformations are invariant under  
 the modified redundant diffeomorphism and the redundant p-form 
symmetry  transformations  of them on-shell.
Thus the symmetries of this model on-shell is interpretable as 
$
{{\rm diffeo} \times {\rm p-form ~sym. } 
\over {\rm modified~ red.~ diffeo.} \times {\rm red.~p-form ~sym. } }.
$
\par
The diffeomorphism transformation is represented by the Lie derivative.
Let ${\cal L}$ denote the Lie derivative and $c$ denotes a
 ghost field of the diffeomorphism~;~
$
{\cal L}_c \Sigma^k =i(c) d\Sigma^k+ d i(c)\Sigma^k
$
where $i(c)$ is the dual operator of the exterior product $\epsilon(c)$ by $c$.
Its adjoint operator is given by 
$
{\cal L}^\ast_c \Sigma^k = \epsilon(c) \delta\Sigma^k
+ \delta \epsilon(c)\Sigma^k,
$
which we  use for the diffeomorphism gauge fixing condition.
$\delta=-\ast d \ast$ is the interior derivative. 
In the above equation, 
$\ast$  denotes the Hodge star dual operation and 
${\cal O}^\ast \equiv - \ast {\cal O} \ast $  is   
the adjoint operator of ${\cal O}$. 
We define the inner products 
of two differential forms 
 $v$ and $w$ $\in ~(2K \oplus O)\otimes \Lambda^i$   by 
$
(v, w)=  tr \int_M \sqrt g   v^{\mu_1 \cdots \mu_i}  w_{\mu_1 \cdots \mu_i} 
\ast 1$, where 
$~~ \ast 1  \cong  \sqrt g dx_1 \wedge dx_2 \wedge dx_3 \wedge dx_4
$
The adjoint operators are defined via
$
(v, D w)=  (D^\ast v, w) 
$
\par 
The transformation of the modified redundant diffeomorphism 
transformation is given by   
$
{\cal L}_\gamma \phi= \Sigma^{\mu \nu} \gamma_\nu,
$
where $\gamma$ is a ghost field of the modified 
redundant diffeomorphism \cite{abe}. 
We take    $ {\cal L}_\gamma^{\ast k} \phi_k = \Sigma^k 
\wedge \phi_k=0 $ as the modified  red.~diffeo. gauge~fix.~conditions. 
\par
The p-form symmetry   for the fundamental field is given 
already. There exist other p-form symmetry and 
the  redundant p-form symmetry   such as     
$\delta_B\ \pi_k=d\Pi_k $~$( \delta_B \ast \pi^i = \delta \ast \pi^i~)
$~ and ~ $\delta_B\phi_k=d\alpha_k $
due to $  d \delta_B \pi^k = d^2 \pi^k = 0$ and $  
d \delta_B\phi^k = d^2 \alpha^k =0$.  
To  fix these symmetries, we set  
$ 
\delta \pi_k=0~{\rm and}~ \delta  \phi_k=0.
$
\par
The gauge fixing conditions of these  
symmetries  or the equations of motion are  summarized as follows, 
where the number of degrees of freedom is given in parentheses~:  
\begin{eqnarray*}
 & {\rm diffeo.~  gauge~fix.~condi.}&~ 
    {\rm red.~diffeo. gauge~fix.~condi. }  \\ \nonumber  
&\check  D_{1B}^\ast \Sigma_k = {\cal L}_c^{k \ast} \Sigma_k = 0 ~(4),    &
\check D_{0 F}^\ast \phi_k= {\cal L}_ \gamma^{ k \ast} \phi_k = 0~(4),
\\ \nonumber 
&  {\rm  p-form.~sym.  /  diff.~ }&
    ~ {\rm   p-form~ sym. /mod.~red.~ diff.~ }   
 \\ \nonumber
&  {\rm  gauge~ fix.~ condi.~or~eq.~of~ mot. }&
    ~ {\rm ~gauge~ fix.~ condi.~or~eq. of~ mot. }   
 \\ \nonumber 
& \hat D_{2B} \Sigma_k = d\Sigma^k=0 ~(9),
         & \hat D_{0 F}^\ast \phi_k= \delta \phi^k=0~(4),  
      \\ \nonumber
&\tilde D_{2B} \Sigma_k ={}^{\rm t.f.} \Sigma^i \wedge \Sigma^j=0 ~(5), 
&\tilde D_{1 F} \phi^i= {}^{\rm t.f.} \Sigma^i \wedge d\phi^j=0 ~(5), 
\\ \nonumber 
 &{\rm p-form~sym. ~gauge ~fix.~ cond.~or~eq.~of~mot. }
& 
\\ \nonumber
& D_{3B}^\ast=  \delta\pi^k=0 ~(3),  & 
\\ \nonumber
&\hat D_{4B}=  d\pi^k=0 ~(9).  & 
\\ \nonumber
\end{eqnarray*}
\par
We use the  decomposition $\Sigma^j = \Sigma^j_0 + \Sigma^j_f$ 
to calculate of the partition function 
where $\Sigma^j_0$ is a back ground solution of the 
 equations of motions and the gauge fixing conditions.
\footnote {We introduce the condition 
${}^{\rm t.f.} \Sigma^i \wedge \Sigma^j=0$ as 
a argument of $\delta-$function 
so $ \Sigma^j_f$ keeps this condition.}
Furthermore we assume that $\Sigma^j_f$  satisfies the linearized 
equation derived from eq. (\ref{eq:three}) so that $\Sigma^j_f$   
represents the deformations     
of the metric and almost complex structures. This  assumption also leads to 
${}^{\rm t.f.}\Sigma^i \wedge d \phi^j=0$.
\par
The quantum action  $S_q$ is given by 
$
S_q = S_0^{\rm red}+ S_{\rm g. f.}~;~
$   
\begin{equation}
S_{\rm g. f.}
= \int_{M_4} \delta_B \{  \bar c  {\cal L}_c^{ \ast ~k}  \Sigma_k{}_f
                       - \bar  \lambda^k   \delta \phi_k  
                       - \bar \gamma  {\cal L}_\gamma^{ \ast ~k}   
                        \phi_k  
                       + \bar \eta^k \delta \pi_k  \} \ast 1 
                       + \delta_B \{ 
                       \pi^{ij}~ {}^{\rm t.f.} \Sigma^i \wedge \Sigma^j  \}, 
\end{equation}
 where $\pi^{ij}$ is a N-L field. 
$\bar c$ and $\bar \gamma$ are anti ghost for the 
diffeomorphism and the redundant diffeomorphism transformation. 
 We are now ready to evaluate the partition function :
\begin{equation}
Z= \int  {\cal D}X (-S_q),
\label{eq:seven}
\end{equation}
where ${\cal D}X$ represents the path integral over the fields 
$ \Sigma^j_f$, ghosts, anti-ghosts and N-L fields. 
In general, these fields contain zero modes and non-zero modes.   
\par
{\bf  The laplacians of the BRST cohomologies }
\par
We introduce the following   deformation complexes. 
The zero modes   are the elements of   
these  cohomology groups of the complexes. 
We can easily check the ellipticity of the deformation complexes.
We may then define the cohomology group.
\begin{equation}
 H^i \equiv {\rm Ker}\, D_i/{\rm Im}\, D_{i-1}= {\rm Ker}\triangle_i.                                                       
\end{equation}
\par 
(i)~ The deformation complex of the bosonic part  :
\begin{eqnarray}
&V_{1B} &=  \Omega^{1, 1}\ni c ,
\\ \nonumber 
&V_{2B} &=  (2K  \oplus O)^\ast  \otimes \Lambda^2 \ni \Sigma^i_f,
\\ \nonumber
&V_{3B} &= \{ (2 K  \oplus 2 K ^{\otimes 2}  \oplus O)^\ast 
\otimes \Lambda^4  \} \oplus  \{(2 K \oplus O)^\ast \otimes \Lambda^3 \} 
\ni (\ast\pi^i, \ast\pi^{ij}),
\\ \nonumber 
&V_{4B} &= (2 K  \oplus O)^\ast  \otimes \Lambda^4 \ni \ast \Pi^i.
\end{eqnarray}
\begin{equation}
  0   \stackrel{D_{0B}} \to    
C^\infty ( V_{1B} ) 
   \stackrel{D_{1B}} \to  
C^\infty (V_{2B}) 
    \stackrel{D_{2B}} \to   
C^\infty (V_{3B})   
    \stackrel{D_{3B}} \to 
C^\infty (V_{4B})   
    \stackrel{D_{4B}} \to  0.
\end{equation}
$D_{0B}$ and $D_{4B}$ are identically zero 
operators.
\begin{eqnarray}
&D_{1B}& : V_{1B} \rightarrow V_{2B},~~
\Sigma^i_f=D_{1B} c 
= \check D_1 c
= {\cal L}_c \Sigma,  
\\ \nonumber
&D_{2B} &:V_{2B} \rightarrow V_{3B },~~
(\ast \pi^i, \ast \pi^{ij})=D_{2B}\Sigma^i_f
=(\hat D_2 \Sigma^i_f, \tilde D_2^i \Sigma^j_f )
=(d\Sigma^i_f,  {}^ {\rm t.f.} \Sigma^i \wedge  \Sigma^j_f),  
\\ \nonumber
&D_{3B}& :V_{3B} \rightarrow V_{4B},~~
\ast \Pi^i= D_{3B} \pi^i= \hat D_3 \ast \pi^i=d \ast \pi^i .
\\ \nonumber
\end{eqnarray}
\par
(ii)~ The  deformation complex of the fermionic part :
\begin{eqnarray}
&V_{0F} & = \{( 2 K\oplus O)^\ast  \otimes \Lambda^0 \} \oplus \Omega^{1, 1}
\ni (\alpha^i, \gamma),~~
\\ \nonumber
&V_{1F}& = ( 2 K\oplus O)^\ast  \otimes \Lambda^1 \ni ( \phi^i)   ,
\\ \nonumber 
&V_{2F} &=\{ (2 K  \oplus 2 K ^{\otimes 2}  \oplus O)^\ast 
\otimes \Lambda^4 \} \oplus \{  (2K  \oplus O)^\ast  \otimes \Lambda^2 \}
\ni (\ast \chi^{ij}) .~~
\end{eqnarray}
\begin{equation}
  0   \stackrel{D_{-1F}} \to    
C^\infty ( V_{0F} ) 
   \stackrel{D_{0F}} \to  
C^\infty (V_{1F}) 
    \stackrel{D_{1F}} \to   
C^\infty (V_{2F})   
    \stackrel{D_{2F}} \to  0.
\end{equation}
$D_{-1F}$ and $D_{2F}$ are identically zero 
operators.
\begin{eqnarray}
&D_{0F}& :V_0 \rightarrow V_1,
~\phi^i =D_{0F}(\alpha^i, \gamma^\nu)
=\hat D_0 \alpha + \check D_0  \gamma
=d\alpha^i+ \gamma^\nu \Sigma_{\mu \nu} dx^\nu,
 \\ \nonumber
&D_{1F}& : V_1 \rightarrow V_2,~~
 \ast \chi^{ij} = D_{1F}\phi^i
=\tilde D_1 \phi^i 
= {}^{\rm t.f.} \Sigma^i \wedge  
d \phi^j, 
\end{eqnarray}
where
$\hat D_i \equiv d~({\rm for}~ i=0 \cdots 3)$, 
$\check D_1 \equiv {\rm Lie~ derivative}$,
$\check D_0 \equiv {\rm modified ~Lie~ derivative}$,
$\tilde D_2 \equiv {}^{\rm t.f.}\Sigma \wedge$, and
$\tilde D_1 \equiv {}^{\rm t.f.}\Sigma \wedge d$.
The laplacians are given by 
\begin{eqnarray}
&(A)~\Delta_{1 B}&=  \check D_1^\ast \check D_1   \mid_{ \Sigma=\Sigma_0}   
\sim \Delta_1 \sim \Delta_{1,0} + \Delta_{0,1}, 
\\ \nonumber
&(B)~\Delta_{2 B}& 
= \{ \check D_1 \check D_1^\ast \oplus  \delta d \} 
\mid_{{}^{t.f.}\Sigma_0^i \wedge \Sigma^j=0}, 
\\ \nonumber 
&(C)~\Delta_{ 3  B}& 
= d \delta  \oplus    \delta  d   \oplus   \tilde D_2 ^{ \ast} 
\tilde D_2 
\sim  \Delta_ {( 2 K\oplus O)^\ast \otimes \Lambda _3} \sim
3\Delta_{1,0}+3\Delta_{0,1}, 
\\ \nonumber 
&(D)~\Delta_{4B}&= d \delta  \sim \Delta_ {( 2 K\oplus O)^\ast 
\otimes \Lambda_4 } \sim 2\Delta_{0,2}+\Delta_{0,0}, 
\end{eqnarray} 
\begin{eqnarray}
&(E)~\Delta_{0 F}&=   \delta d \oplus  \check D_0 ^\ast \check D_0 
\mid_{\Sigma=\Sigma_0}
\sim  \Delta_ {( 2 K\oplus O)^\ast \otimes \Lambda_0}
\sim 2\Delta_{0,2}+\Delta_{0,0}, 
\\ \nonumber
&(F)~\Delta_{1 F}&  
= \{ d \delta  + \check D_0 \check D_0^\ast \} 
\mid_ {{}^{\rm t.f.} \Sigma_0^i \wedge  d  \phi^j=0}, 
\\ \nonumber 
&(G)~\Delta_{2 F}& 
=\tilde D_1 \tilde D_1^\ast  \sim  
  \Delta_ {(2 K  \oplus 2 K ^{\otimes 2}  \oplus O)^\ast  \otimes 
\Lambda_4 }
\sim 2\Delta_{2,0}+3\Delta_{0,0}
\end{eqnarray} 

where $\triangle_i = D_{i-1}^{} D_{i-1}^* + D_i^* D_i^{}$ 
and  $D_1^\ast D_1 \cong \hat D_1^\ast \hat  D_1  \oplus 
\tilde D_1^\ast \tilde  D_1$ etc. 
These laplacians contain only the back ground field 
 $\Sigma_0$. For example, 
the adjoint operator of $\tilde D_{1F}$ and  $\tilde D_{2B}$ are 
given by  
$
\tilde D_{1 F}^{\ast, i} \sigma^{(ij)}_{[\mu \nu \rho \tau]} 
= {\rm tr}_i (\Sigma_0^i d)^{ ~ \mu \nu \rho} 
\sigma^{(ij)}_{[\mu \nu \tau \rho]},~  
\tilde D_{2 B}^{\ast, i} \sigma^{(ij)}_{[\mu \nu \rho \tau]} 
= {\rm tr}_i \Sigma_0^{i ~ \mu \nu} \sigma^{(ij)}_{[\mu \nu \tau \rho]}  
$.
Some useful  expressions  about  $\Sigma_0^k$ 
  are given by 
\begin{equation}
\Sigma_0{} ^k_{\mu \nu} \Sigma_0^{k\tau \rho}= 2P_{\mu \nu}{} ^{\rho \tau}
 \equiv 2(\delta^{ [\tau }_\mu \delta _\nu^{\rho ]}-{1 \over 2} 
\epsilon_{\mu \nu}{}^{\tau \rho} )
\Sigma_0{}^{k \rho }_{\mu } \Sigma_0 {}^{l \nu}_{ \rho}= 
\epsilon _{k l n } \Sigma_{ 0 \mu} ^{ n \nu}.
\end{equation}
$\Delta_i$ denotes the de Rham laplacian  which operates on i-form. 
We explain how to obtain  each laplacians in short. 
\par
(A)~;~
  $\Delta_{1B}$   reduces to the 
de Rham laplacian $\Delta_1$ by using the killing equation
and  the Rich flatness 
 via  $g_{\mu \nu}(\Sigma_0)$.
\par
(C), (E) ~;~
In these cases, zero modes of $\Delta_{3B} ( \Delta_{0F}) $ 
are direct sum of  two parts  
$\Delta_ {( 2 K\oplus O)^\ast \otimes \Lambda _3}
( \Delta_ {( 2 K\oplus O)^\ast \otimes \Lambda_0} )$
and $\tilde D_2^\ast \tilde D_2 ( \check D_0^\ast \check D_0 )$.
~ $\tilde D_2^\ast \tilde D_2$ and $\check D_0^\ast \check D_0)$ 
become constant number by the property of  $\Sigma_0$. 
So their zero modes  are 0 and  do not contribute  
to the zero modes of the laplacians. 
$\Delta_{3B}$ and $\Delta_{0F}$ reduce to the Dolbeault laplacians.  
\par
(G)~;~
$\tilde D_1\tilde D_1^\ast$ reduces to $\Delta_
{(2K \oplus 2K^{\otimes 2} \oplus O)^\ast \otimes \Lambda_4}$ by substituting 
$\Sigma_0$.  Their  zero modes are   $(2K \oplus 2K^{\otimes 2} \oplus O)^\ast$ bundle valued harmonic 4-forms.
\par
(D)~;~
$\Delta_{4B}$  is  $\Delta_{(2K \oplus O)^\ast \otimes \Lambda_ 4}$.   
Their  zero modes are   $(2K \oplus  O)^\ast$ bundle  
valued harmonic 4-forms.
\par
(B),~(F)~;~
 In these cases, the situations are more complicated. 
From the gauge fixing conditions  and the equations of motion, 
 $\delta\Sigma_0^k=0$ and $d\phi^k=0$  can be derived 
in addition to $d\Sigma^k=\delta\phi^k=0$. 
So the zero modes of  $\Delta_{2B}$ and $\Delta_{1F}$  are 
at least the zero modes of $\Delta_i$. 
We cannot derive the explicit form of $\Delta_{2B}$ and $\Delta_{1F}$ 
 since  ${}^{\rm t.f.}\Sigma^k \wedge 
\Sigma^i=0$  and ${}^{\rm t.f.}\Sigma^k \wedge d\phi^i=0$  restrict 
the zero modes in the complicated way.  
However, we can discuss about the Seeley's coefficients 
for $\Delta_{2B}$  and $\Delta_{1F}$ later. 
\par
{\bf  The dimensions of the moduli spaces} 
\par
 The dimensions  of $H^i$   are finite and represented by $h^i$.
 $H_2{}_B$ ($H_1{}_F$) is exactly identical with  ${\cal T( M }(\Sigma) )$ 
(${\cal T( M }(\phi) )$) which is 
the tangent space  of moduli space  
${\cal M}(\Sigma)$  (${\cal M}(\phi)$) :
\begin{equation}
  T({\cal M}(\Sigma)) 
  = \{  \Sigma^k_f \vert \Sigma^k_f \in (2 K \oplus O)^\ast  \otimes 
\wedge^2,  D_{2B}   \Sigma^k_f = 0 \} / 
              diffeo.   \ .          
                                                  \label{eq:twentytwo}
\end{equation}
\begin{equation}
  T({\cal M}(\phi)) 
  = \{  \phi^k \vert \phi^k \in (2 K \oplus O)^\ast  \otimes 
\wedge^1,  D_{1F}   \phi^k = 0 \} /      mod.~ red.~diffeo.    .          
\end{equation}
\par
The dimensions of the moduli spaces of 
 ${\cal M}(\Sigma)$ \cite{abe} and 
 (${\cal M}(\phi)$)\footnote  {We thank T.  Ueno since he 
did some similar calculations about the fermionic moduli.}  
   are given by as follows  
by  the Atiyah-Singer index theorem\cite{shanahan} , 
\begin{eqnarray}
 {\rm Index}  {\rm~ of~ eq.}~(19) &=&  \Sigma^3_{i=0} (-1)^i h^i_B
\\ \nonumber 
&=&\int_{M_4} \frac{ {\rm td} (TM_4 \otimes {\bf C} ) }
      {{\rm e} (TM_4) } \cdot  {\rm ch} \{ \sum_{n=0}^3 \oplus (-1)^n V_{nB} \} 
\\ \nonumber
&=& 2 \chi + 7 \tau \rightarrow 2 \chi - 7 \mid \tau \mid \, , 
                                             \label{eq:twentyfour}
\end{eqnarray} 
\begin{eqnarray}
 {\rm Index}{\rm~ of~ eq.}~(22)&=&  \Sigma^3_{i=0} (-1)^i h^i_F
\\ \nonumber 
&=&\int_{M_4} \frac{ {\rm td} (TM_4 \otimes {\bf C} ) }
      {{\rm e} (TM_4) } \cdot  {\rm ch} \{ \sum_{n=0}^3 \oplus (-1)^n V_{n F} \} 
\\ \nonumber
&=& 5 \chi + 7 \tau \rightarrow 5 \chi - 7 \mid \tau \mid \, , 
                                             \label{eq:twentyfive}
\end{eqnarray} 
where ${\rm ch},$ ${\rm e}$ and ${\rm td}$ are the Chern 
character, Euler class and Todd class of the various vector bundles 
involved. 
The index is determined by the Euler number $\chi=\int_{M_4}
x_1 x_2$ and  Hirzebruch signature $\tau=\int_{M_4}{x_1^2+x_2^2 \over 3} $.
$x_i$ denotes the first Chern classes of $L_i$ or $\bar {L_i}$. 
The dimension of the fermionic moduli space of this model 
is zero  on $K3$ and non-zero on $T^4$
  while that  of the   bosonic   moduli space is  not zero 
 on $K3$ and $T^4$.   
\begin{eqnarray}
&{\rm dim.}~{\cal  M}(\Sigma)
&=h_B^2=2\chi- 7\mid \tau \mid +h_B^1+h_B^3-h_B^4, 
\\ \nonumber
&~~~~~~&{\rm with~ containing~ the ~scale~ factor},
\\ \nonumber
&{\rm dim.}~{\cal  M}(\phi)
&=h_F^1=5\chi- 7\mid \tau \mid  +h_F^0+h_F^2.
\end{eqnarray}
\begin{equation}
h^1_B=b_1,~h^3_B=3b_1,~h^4_B=3b_0,~h^0_F=3b_0,~h^2_F=5b_0,
\end{equation}
where $b_i$ is the i-th Betti number.
The result is summarized as follows;
\begin{center}
\begin{tabular}{|l|l|l|l|l|l|l|r|} \hline
\it       &dim. $ {\cal M} (\Sigma) $ &dim. $ {\cal M}(\phi)  $ 
&$\chi$    &$\tau $  &$b_0$ &$b_1 $&$b_2$\\ 
\hline
\it $K_3$ &61  & 0  &24     &-16    &  1 & 0 & 20       \\ 
\hline
\it $T^4$ &  13  &  8  & 0   & 0 & 1 & 4& 6      \\
\hline
\end{tabular} 
\end{center}
Thus the bosonic moduli spaces agree with those of the Witten type model 
but the fermionic moduli spaces do not agree.
\\
{\bf  The partition function  up to  one-loop corrections}
\par
When expanded out   by using the properties of $\delta_B$ 
\cite{abe},   the quantum action is given  by 
\begin{eqnarray}
S_q & =& 
B \ast T B^t -F \ast \tilde T F^t
\cr
&+& \bar \gamma \ast  \tilde \Delta_{0F} \gamma
- \bar c \ast \Delta_{1B} c
\cr
&+& \bar \eta^k \ast \Delta _{4B} \eta_k + 
\bar \lambda \ast \hat \Delta_{0 F} \lambda + 
{\rm other~ higher~ order~ terms},   
\end{eqnarray}
with some field redefinition.
We integrate over non-zero modes.
The Gaussian integrals over the commuting 
$  \bar \gamma -  \gamma$~ and $\bar \lambda^k-\lambda_k $
sets of fields give   
$
\det (\Delta_{0 F } ) ^{-1}.  
$
While the anti-commuting sets of   $  \bar c $ -  $c$ and 
$\bar \eta^k- \eta_k$~ do
$
\det (\Delta_{1B} ) \det (\Delta_{4 B} ).
$
We integrate  over the remaining   
$ B \equiv (\pi_c,~~ \Sigma^k_f,~~\pi^k \oplus \pi^{ij},~~
(\pi_{\lambda})^k ) $
-system  and 
$F \equiv \{ \chi_{\gamma},~~\phi^k,~~\chi_{ij}  \}$- system
 by taking ${\rm det}T ={\rm det}^{1 \over 2} (T^\ast T)$ 
(${\rm det} \tilde T ={\rm det}^{1 \over 2} ( \tilde T^\ast  \tilde T)$ )
 \cite{blau} and  using the nilpotency $D_iD_{i+1}=0$ . 
\begin{equation}
\det (T^\ast T)^{1 \over 2}  
= \{ \Pi_{i=1}^4 \det(\Delta_{iB})
 \}^{1 \over 4},~~
\det \tilde (T^\ast \tilde T)^{1 \over 2}  
=\{ \Pi_{j=0}^2 \det (\Delta_{j F})\}^{1 \over 4}.
\end{equation}
The bosonic N-L fields are represented by 
$\pi$   and   the fermionic ones are by $\chi$. 
Their contributions do not cancel out each other. 
\par
The partition function on $K_3$  leads to 
\begin{equation}
Z=  \int  d({\rm zero~ modes}) \Sigma^\prime 
\big[  
{ 
\Pi_{j=0}^2{\rm det} \Delta_{jF}
\over 
\Pi_{i=0}^4{\rm det} \Delta_{iB}
} 
\big]
^{1 \over 4}
\cdot 
{ 
\{ 
({\rm det} \Delta_{1 B}) 
({\rm det} \Delta_{4 B}) 
\} 
\over
\{ 
({\rm det} \Delta_{0 F}) 
\}  
}.
\end{equation}
\par
{\bf The discussion of the integrand of the partition function}
\par
By the results of the index theorem in eq. (22) and (23), 
we can derive the Seeley's coefficients for 
$\Delta_{1F}$ and $\Delta_{2B}$ for on-shell
 by substituting  
  the other contributions of  deRham Laplacians \cite{gilkey} and 
the Dolbeault laplacians. 
The index theorem gives as follows; 
 \begin{eqnarray}
\int  \Sigma_{i=1}^4 (-1)^i  a_l( \Delta _{iB})& = -{8 \over 3}\chi
\mid_{\rm on-shell}  
\\ 
\nonumber
\int  \Sigma_{j=0}^2 (-1)^j  a_l( \Delta _{jF}) &= -{1 \over 3}\chi
\mid_{\rm on-shell}  
\end{eqnarray}
where $a_l ~(l=0,2,4)$ represents the Seeley's coefficient.
The method of the $\zeta$-functional 
regularization is reviewed  in the reference  \cite{blau}.  
We follow the notations of Gilkey \cite{gilkey} for the 
Seeley's coefficients.  
From the consideration  before and ref. \cite{abe}
about the laplacians $\Delta_{1F}$ and $\Delta_{2B}$,
 we can expect that 
\begin{equation}
\Delta_{2B} \mid_{\rm on-shell} =  \Delta_
{  O \otimes 3\Lambda_{1,1} }+
\Delta_{K^\ast \otimes \Lambda_{0,2}}
 \sim 3\Delta_{1,1}+\Delta_{0,0}, 
\end{equation}
\begin{equation}
\Delta_{1F} \mid_{\rm on-shell} =  \Delta_ {(K\oplus O)^\ast \otimes \Lambda_1}\sim 2\Delta_{0,1}+2\Delta_{1,0},
\end{equation}
where $\Delta_{i,j}$ represents the Dolbeault laplacians on $(i,j)$ form.
The Dolbeault laplacians are defined as follows :
The exterior differential spilts as  
  \begin{equation}
d=d^\prime + d^{\prime \prime}
\end{equation}
where 
\begin{equation}
d^\prime : {\cal D}^{p,q} \rightarrow  {\cal D}^{p+1,q},
 \ \  d^{\prime \prime} : {\cal D}^{p,q} \rightarrow  {\cal D}^{p,q+1}.
\end{equation}
${\cal D}_{p,q}(M)$ denotes the space of of $C^\infty$ complex 
(p,q) forms on M.
 Their adjoint operators are  
\begin{equation}
\delta^\prime =-\ast d^\prime \ast : 
 {\cal D}^{p,q} \rightarrow  {\cal D}^{p-1,q},
 \ \ \delta^{\prime \prime} =-\ast d^{\prime \prime}\ast :
 {\cal D}^{p,q} \rightarrow  {\cal D}^{p,q-1}.
\end{equation}
Thus 
$\Delta_{p,q}$ is given by 
\begin{equation}
\Delta_{p,q}=-(\delta^{\prime \prime} d^{\prime \prime}+
d^{\prime \prime} \delta^{\prime \prime}): 
 {\cal D}^{p,q} \rightarrow  {\cal D}^{p,q}.
\end{equation}
\par
The Seeley's coefficients for $\Delta_{1F}$ and $\Delta_{2B}$ 
which are derived by the index theorem agree with 
those of the expected laplacians in eq. (29) on-shell
\footnote{ On-shell conditions mean Riemannian half-flat  
 so $ \tau \mid_{\rm on-shell} ={2 \over 3}\chi\mid_{\rm on-shell} 
 = \int (4\pi)^{-2} {1 \over 3} R_{\mu \nu \rho \sigma}^2$.}.
If we use $\Delta_{1F} \mid_{\rm on-shell} = 2 \Delta_1$ , 
we can deform $Z$ as follows. 
\begin{equation}
Z \mid_{\rm on-shell} = \int d({\rm  zero~ modes} )
\big[  
{ 
\Pi_{j=0}^2{\rm det} \Delta_{jF} ^{(-1)^j}
\over 
\Pi_{i=1}^4{\rm det} \Delta_{iB} ^{(-1)^i}
} 
\big]
^{1 \over 4} .
\end{equation}  
Furthermore, if we use both equations for $\Delta_{2B}$
and $\Delta_{1F}$, 
then we can show that the physical 
degrees of the freedom of $Z$ in the  field 
theoretic sense \cite{blau} is zero, as expected from the property of the 
elliptic complex. Thus the phase space is finite and 
only zero modes work.
\par
Now we compare the Abelian BF-type model and the diffeo. BF-type model.
The situation as there  are no degrees of the freedom in the 
partition function is similar to that of the partition function of the 
Abelian BF- type model given by Schwarz \cite{schwarz,blau}.  
The result of the path-integral over the non-zero modes 
  of the partition function (i.e., the integrand of the zero modes)  
 for the Abelian BF-type model is given by the ratio of some 
determinants of the de Rham laplacians. 
(On-shell condition in this case means that the Abelian gauge field  and 
the  two-form field  are both flat.) 
\cite{blau}. 
The ratio of these determinants  
 is represented by the  $d=4$ R-torsion  
(i.e., some topological invariant).  It  
becomes trivial  for the compact even dimensional manifolds  
without boundaries  when the cohomologies  
  are  trivial  \cite{blau,ray}. 
\par In our case we show that the integrand of the 
partition function is represented by some $\bar \partial $-torsions. 
 Both of $\Pi_{j=0}^2{\rm det} \Delta_{jF} ^{(-1)^j}$ 
and $\Pi_{i=1}^4{\rm det} \Delta_{iB} ^{(-1)^i}$ are reduced to as follows. 
\begin{equation}
Z=\int \big[
{(\Pi_{i=0}^2 {\rm det}\Delta_{0,i}^{(-1)^i})^2
(\Pi_{i=0}^2{\rm det} \Delta_{i,0}^{(-1)^i})^2
\over
(\Pi_{i=0}^2 {\rm det}\Delta_{i,1}^{(-1)^i})^{-1}
(\Pi_{i=0}^2 {\rm det}\Delta_{1,i}^{(-1)^i})^{-2}
(\Pi_{i=0}^2 {\rm det}\Delta_{0,i}^{(-1)^i})^2
}\big]^{1\over 4}.
\end{equation}
We introduce the definition of the $\bar \partial $-torsion by using the 
zeta function. 
The zeta function associated with the Laplacian 
$\Delta_{p,q}$ is defined by 
\begin{equation}
\zeta_{p,q}={1\over \Gamma(s)} \int ^\infty_0t^{s-1}\rm{tr}(e^{t\Delta_{p,q}} -
P_{p,q}) dt = \Sigma_{\lambda_n < 0} (-\lambda_n)^{-s},
\end{equation}
for Re s large, the sum running over the non-zero eigenvalues $\lambda_n$ of 
$\Delta_{p,q}$.~ $P_{p,q}$ is the projections of ${\cal D}^{p,q}$ onto the subspace of harmonic forms in ${\cal D}^{p,q}$.
\par
The definition of the $\bar \partial $-torsion is given by as follows : 
Let M be a compact complex analytic manifold without boundary, of complex dimension N, and let $\chi$ be a finite dimensional unitary 
representation of the fundamental group $\pi_1(M)$.
Suppose a Hermitian metric is defined on M. For each integer 
$p=0, \cdots N$, the $\bar \partial $-torsion is defined 
as the positive root of 
\begin{equation}
\log {\rm   T_p} (M, \chi)
= {1\over 2} \Sigma^N_{q=0} (-1)^q q \zeta^\prime_{p,q}(0,\chi),
\end{equation}
where the zeta function $\zeta_{p,q}$ is defined by the above equation.
\par 
The invariance theorem about $\bar \partial $-torsion 
\cite{ray} is that  the ratio   $T_P(M,\chi_1) /T_P(M,\chi_2)$  
 is metric independent  when M  keeps some conditions. 
$\chi_1$ and $\chi_2$ are two 
representations of the fundamental group $\pi_1(M)$.
\par
The another useful theorem for $\bar \partial$-torsion 
is as follows\cite{ray}:
Let M be a closed Kahler manifold of complex dimension N, and 
$\chi$ a finite dimensional unitary representation of the fundamental group 
$\pi_1(M)$. The $\bar \partial $-torsion of M, 
defined for a Kahler metric, satisfies 
\begin{equation}
\Sigma^N_{p=0}(-1)^p {\rm log} T_p(M,\chi)=0.
\end{equation}
This is because the above alternative sum  becomes 
 the R-torsion and vanishes since the real dimension 
of M is even.
\par
By using above theorems, the partition function becomes 
\begin{equation}
Z= \int [T_0^\ast(M,\chi) T_0(M,\chi)]^2,  
\end{equation}
where $\chi$ is trivial. 
For the flat torus case, there is 
the theorem \cite{ray}that  $T_p( M, \chi)=1$ 
for the complex dimension $ N\ {\rm of} \ M > 1 $ case . 
Thus for $T_4$, the integrand of the partition function is trivial.
For K3, the fundamental group is trivial and  
the integrand of the partition function 
has the modular dependence. This result can be expected 
since the model is the topological gravity model 
and the integral of the partition function is 
over the moduli.   
In both cases, the integrand of the partition function has no divergence 
since all $\Delta_{p,q}$ has no contribution of  zero modes.
\par 
The partition function on $T^4$ is zero 
after the path-integral over the fermionic moduli $\phi$ 
since dim. ${\cal M}(\phi) \not= 0$.
While on K3,   the partition function does not vanish
since the dimension of the fermionic moduli is zero. 
Our attempt is the extension of the Schwarz's model in  Riemannian 
manifolds to  the  hyperkahler manifolds.
If we can take the other gauge fixing conditions 
for the fermionic moduli, then 
these situations  would be changed. To  modify the moduli spaces 
to relate the compactified string theory, the addition of the fundamental 
field would be necessary.
\newpage
{\bf Acknowledgment}
\par
 We are grateful to  T. Ueno, A. Nakamichi,   
Q-Han Park, S. Morita and  N. Sakai 
for useful discussions.
We  thank  A. Futaki  most for pointing out  
   the difference between $K(g)$  and ${\rm M} (\Sigma)$ and the 
necessity of  an  almost complex structure with  vanishing real 
first Chern class for the $\Lambda =0$ case.  
\newpage

\end{document}